\documentclass[twocolumn,superscriptaddress,nofootinbib,preprintnumbers,showpacs,tightenlines,notitlepage]{revtex4}
\usepackage[latin1]{inputenc}
\usepackage{graphicx}
\usepackage{amssymb}
\usepackage{comment}
\usepackage{float}
\usepackage{amsmath}
\usepackage{amsfonts}
\usepackage{dcolumn}
\usepackage{hyperref}
\usepackage{amsthm}
\usepackage{enumerate}
\usepackage{bm}

\def\udot{\dot{u}}
\newcommand{\w}[1]{\bm{#1}} %bold greek
\def\ud{\textrm{d}}

\def\be {\begin{equation}}
\def\ee {\end{equation}}
\def\ba {\begin{eqnarray}}
\def\ea {\end{eqnarray}}
\def\vl{\w{\ell}}
\def\vk{\w{k}}
\def\vm{\w{m}}
\def\vmbar{\w{\overline{m}}}

\newcommand{\smfrac}[2]{{\textstyle{\frac{#1}{#2}}}}

\begin{document}

\title{Rotating and twisting locally rotationally symmetric imperfect fluids}

\author{Norbert Van den Bergh}
\email{norbert.vandenbergh@ugent.be}
\affiliation{Ghent University, Department of Mathematical Analysis EA16, \\ Krijgslaan 281, 9000 Ghent, Belgium.}
\begin{abstract}
 Rotating and twisting locally rotationally symmetric imperfect fluids in general relativity admit a much larger set of
 solutions than the self-similar ones recently suggested in the literature. Explicit forms of the metrics are given and the self-similar sub-cases are discussed.
\end{abstract}
\maketitle

\section{Introduction}

{\em Locally Isotropic} space-times are space-times possessing at each point $P$ a continuous isotropy group $I_P$, which must be a subgroup of the Lorentz group \cite{CahenDefrise_1968, Kramer}.
More precisely  a space-time is said to be {\em Locally Rotationally Symmetric} (LRS) \cite{Ellis_1967,Stewart_Ellis_1968} if, in addition
to the previous property, there exists a time-like vector field $\w{u}$,  such that $\w{u}_P$ is invariant under the action of $I_P$ at each point $P$. It follows that $I_P$ acts in the subspace $V_P$ orthogonal to $\w{u}_P$ and hence is a
one- or three-dimensional group of rotations. As the three-dimensional case leads to the Friedmann-Lema\^{i}tre-Robertson-Walker (FLRW) models, one usually restricts to the case of a one-parameter group.
For any given vector field $\w{u}$, invariant under the group, there is
then at each point $P$ a unique space-like unit vector $\w{v}_P$, defining the {\em local axis of rotation}, orthogonal to $\w{u}_P$ in $V_P$.
When $\w{u}$ is the velocity field of a perfect fluid, $\w{u}$ is the unique time-like eigenvector of the Ricci tensor and is hereby automatically
invariant under the action of the isometry group: the corresponding LRS spacetimes have been completely
analysed by Ellis \cite{Ellis_1967} and Stewart and Ellis \cite{Stewart_Ellis_1968}. It was proved that for perfect fluid spacetimes the vorticity $\omega$ of $\w{u}$ and the twist $\hat{\omega}$ of $\w{v}$ necessarily have a vanishing product,
$\omega \hat{\omega}=0$, implying the
division of the corresponding spacetimes in three distinct classes:
LRS-I ($\omega \neq 0 =\hat{\omega}$), LRS-II ($\omega = 0 =\hat{\omega}$) and LRS-III ($\omega = 0 \neq \hat{\omega}$).
LRS-II spacetimes have been extensively studied and have found wide applications in the study of spherically symmetric objects. LRS-I space-times are stationary, non-expanding and shear-free generalizations of the G\"odel metric, while
LRS-III spacetimes are non-rotating and geodesic. As this restricts somewhat the application of local rotational symmetry to the study of rotating matter, an attempt was made recently by Singh
et al. \cite{Singh2016, Singh2017,Singh2017b} to relax the condition of a perfect fluid and consider in stead an imperfect fluid with a flow vector assumed to be invariant under the isometry group.
For this purpose a 1+1+2 covariant formalism was used, whereby the usual kinematic quantities of the flow together with the Ricci and Weyl tensors were decomposed with respect to the time-like and
space-like unit vectors $\w{u}$ and $\w{v}$. Local rotational symmetry was then imposed by equating to zero all vectorial and tensorial quantities in
the orthogonal complement of the
$(\w{u},\w{v})$ plane. This leaves one with a set
of scalars, invariantly defined by $\w{u}$, for which a reduced set of field equations could be constructed. These scalars comprise: the expansion $\Theta$ of $\w{u}$, the
projections
$\udot, \Omega, Q$ on $\w{v}$ of $\w{u}$'s acceleration and
vorticity and of the heat flow, the simple eigenvalues  $\Sigma, {\cal E}, {\cal H}, \Pi$ of the shear, the electric and magnetic parts of the Weyl tensor and of the anisotropic stress, the matter density $\mu$ and the
pressure $ p$, the twist $ \xi$ of $\w{v}$ and the expansion $\phi$ of the local two-sheets orthogonal to the $(\w{u},\w{v})$ plane (for details see \cite{Singh2016} and section 2 below).
A key result obtained in this way is relation (74) of \cite{Singh2016}, which says that for any scalar $\Psi$
\begin{equation}\label{keyeq}
 \Omega \w{u}(\Psi) = \xi \w{v}(\Psi).
\end{equation}
From this equation the authors claimed that spacetimes with $\Omega \xi \neq 0$ ought to be self-similar, after which they proceeded with a further reduction of the field equations to a system of ODE's, the consistency of which was
proved, however without giving the explicit metric forms. The given explanation for the self-similarity was `that (\ref{keyeq}) is unchanged under the transformation  $\tau\rightarrow a\tau$,
$\rho \rightarrow a\rho,$ where $\tau$ and $\rho$ are the curve parameters of the integral curves of $\w{u}$ and $\w{v}$ respectively'. Aside from the fact that it is not legitimate to write $\w{u}=
\frac{\partial}{\partial \tau}$ and $\w{v}=
\frac{\partial}{\partial \rho}$ for the (generally) non-commuting vector fields $\w{u}$ and $\w{v}$, this conclusion is certainly premature: choosing an orthonormal basis $(\w{e_0}=\w{u}, \w{e_1}=\w{v},\w{e_2},\w{e_3})$ with dual basis
$(\w{\theta^0}, \w{\theta^1}, \w{\theta^2}, \w{\theta^3})$,
the only information one can infer from (\ref{keyeq}) is that for any scalar $\Psi$ (hence necessarily satisfying $\w{e_2}(\Psi)=\w{e_3}(\Psi)=0$),
\begin{equation}\label{key2}
 \ud \Psi = \w{e_0}(\Psi) \w{\theta^0} + \w{e_1}(\Psi) \w{\theta^1} = \Omega^{-1}\w{e_1}(\Psi)( \xi \w{\theta^0}+ \Omega \w{\theta^1} ),
\end{equation}
such that %, provided $\Omega \xi \neq 0$,
all scalars are functionally dependent. This, of course, should not be surprising \cite{CahenDefrise_1968}, as all LRS spacetimes have either a 4 dimensional group of isometries
acting multiply transitively on orbits which are 3 dimensional sub-manifolds
(and hence they are hypersurface-homogeneous) or have a 3 dimensional group of isometries acting multiply transitively on 2 dimensional orbits (corresponding to the LRS-II case, in which any vector field orthogonal to the orbits has vanishing
vorticity $\Omega$).

In order to construct the most general locally rotationally symmetric imperfect fluid, we notice \cite{Kramer} that only Petrov types D or O admit a continuous isotropy group of spatial
rotations, a property which in \cite{Singh2016} is also clear from the fact that the electric and magnetic components of the Weyl tensor take the form $\w{E}=\textrm{diag}(0,\mathcal{E},-\mathcal{E}/2,-\mathcal{E}/2)$ and
$\w{H}=\textrm{diag}(0,\mathcal{H},-\mathcal{H}/2,-\mathcal{H}/2)$.
Omitting the conformally flat case, the $(\w{e_0},\w{e_1})$ plane considered in \cite{Singh2016} is thus uniquely fixed by the type D condition, namely as the plane spanned by the real principal null directions of the Weyl
tensor. This has lead Goode and
Wainwright (GW) \cite{GoodeWainwright_1986} to give an alternative characterisation of LRS spacetimes, based on the canonical form of the Weyl tensor for Petrov type D. On the basis of this characterisation
(which runs parallel to the
one given by Stewart and Ellis in \cite{Stewart_Ellis_1968}) they show that non-conformally flat LRS spacetimes of Petrov type D necessarily belong to one of four mutually disjoint classes\footnote{class
A is LRS-II, while B1 and B2 coincide with LRS-I and LRS-III respectively; B3 (which also appears in \cite{Petrov1969}, pp.~232, 233) was omitted for no obvious reason from the discussion in \cite{Stewart_Ellis_1968}} (A, B1--3).
For each of these classes the
general form of the metric was constructed, together with a canonical null tetrad $(\tilde{\vk},\tilde{\vl}, \tilde{\vm}, \tilde{\vmbar})$, with the pair $(\tilde{\vm}, \tilde{\vmbar})$ being defined up to constant rotations and the
pair $(\tilde{\vk},\tilde{\vl})$
being fixed up to discrete transformations, at least when the spin coefficients $\tilde{\rho}$ and $\tilde{\mu}$ are not both
0. The latter situation occurs as a sub-case of class A, in which both $\tilde{\rho}$ and $\tilde{\mu}$ are real: as this implies that any
vector field in the $(\tilde{\vk},\tilde{\vl})$ plane has a vanishing vorticity, this will not be considered further.

It follows that {\em any} (non-conformally flat) LRS spacetime of Petrov type D, with an energy-momentum tensor of a general imperfect fluid (determined by a rotating flow which is invariant under the isometry group)
is necessarily described by a member of the GW classes B1--3. Furthermore the invariant vector fields $\w{u},\w{v}$ appearing in \cite{Singh2016} are then related to the canonical GW null vectors $\tilde{\vk},\tilde{\vl}$ by
\be \w{u}=\w{e_0} =\frac{\vk+\vl}{\sqrt{2}},\
\w{v}=\w{e_1}=\frac{\vk-\vl}{\sqrt{2}}, \label{e0e1def}
\ee
and a boost
\be \label{boosts}
\tilde{\vk}=A^{-1} \vk,\  \tilde{\vl}= A \vl,
\ee
with $A$ a scalar which is invariant under the isometry group, but which is otherwise a completely arbitrary function.

In section 2 I give a review of the GW construction
%, comparing their subsequent tetrad fixations with the ones made in \cite{Stewart_Ellis_1968}.
and I give explicit expressions for the kinematic quantities %and the components of the energy-momentum tensor
of an imperfect fluid, obtained by boosting the canonical tetrad. It follows that {\em any} of the B1--3 spacetimes can be interpreted as a locally rotationally symmetric imperfect fluid, a conclusion which
also invalidates the results in \cite{Singh2017b}, where the authors again use the supposed self-similarity to conclude that $f(R)$ theories of gravity (which obviously only will allow self-similar solutions when $f(R)\sim R$)
cannot admit rotating and twisting imperfect fluids in the presence of an LRS background.
In section 3 I have a closer look at the self-similar cases
considered by Singh et al.~and show that self-similar solutions do indeed occur in each of the classes B1, B2 or B3. The possible metrics are given explicitly, together with the explicit expressions for the matter variables.

Notations and sign conventions are as in \cite{GoodeWainwright_1986, Kramer} and some familiarity with the orthonormal tetrad and Newman-Penrose formalisms (see for example \cite{MacCallum_Cargese} and \cite{PenroseRindler}) is assumed.

\section{Construction of the canonical tetrad}
It is clear that the LRS property, after choosing a Newman-Penrose null tetrad
$(\vk,\vl,\vm,\vmbar)$
such that $\Psi_2$ is the only non-vanishing component of the Weyl spinor (fixing $\vk,\vl$ up to a boost and $\vm, \vmbar$ up to a spatial rotation), implies the GW conditions\footnote{the reason being that otherwise any
of these coefficients, each having non-vanishing spin-weight, can be used to break the rotation isotropy}
\ba \label{GW_main}
 C1 : & \kappa=\sigma=\tau=\lambda=\nu=\pi=0,\\
 C2 : & \phi_{01}=\phi_{12}=0,\\
 C3 : & \delta R =0.
\ea
The essential content of the GW paper is that also the reverse holds, namely that conditions C1-C3 enforce complete isotropy in the $(\w{m},\w{\overline{m}})$ plane. In a nutshell the proof of the theorem proceeds as follows:
first the Newman-Penrose equations and C1-C2 imply $\phi_{02}=0$, after which a boost is shown to exist, making
\be \alpha+\overline{\beta}=0.\ee

Introducing an orthonormal tetrad\footnote{see the appendix of \cite{GoodeWainwright_1986} for a conversion between Newman-Penrose and orthonormal tetrad variables}
with $\w{e_0},\w{e_1}$ defined by (\ref{e0e1def}), with
\be \label{mmbardef}
\w{e_2} =\smfrac{1}{\sqrt{2}} (\vm+\vmbar), \ \w{e_3}=\smfrac{1}{\sqrt{2}i} (\vm-\vmbar)
\ee
and with an energy-momentum tensor
\be \label{energymomentum}
\w{T}= (\mu+p) \w{u}\otimes \w{u} +p \w{g} + \w{q}\otimes \w{u} +\w{u}\otimes \w{q} +\w{\pi}
\ee
(flow vector $\w{u}$, heat flow $\w{q}$ and anisotropic stress $\w{\pi}$),
this implies for the kinematic quantities of $\w{u}=\w{e_0}$ the conditions
\ba
\omega_2 &=& \omega_3=0,\\
\dot{u}_2 &=& \dot{u}_3=0,\\
\sigma_{22}-\sigma_{33} &=& \sigma_{12}=\sigma_{13}=\sigma_{23}=0,\\
q_2 &=&q_3=\pi_{12}=\pi_{13}=\pi_{23}=0.
\ea
The Kundt-Sch\"ucking-Behr variables \cite{EllisMac} associated to the orthonormal tetrad obey then
\ba
n_{22}-n_{33} &=& n_{12}+a_3=n_{13}-a_2=n_{23}=0,\\
\Omega_2 &=& \Omega_3=0.
\ea
The vorticity of $\w{u}$ can therefore be written as $\w{\omega}=\omega_1 \w{e_1}$, with
\be \label{omega1rhomu}
\omega_1=-\smfrac{i}{2\sqrt{2}}[(\mu-\overline{\mu})+(\rho-\overline{\rho})],
\ee
while a calculation of the kinematic quantities of the spatial vector field $\w{e_1}$ \cite{GoodeWainwright_1986} shows that its
twist
%\footnote{in \cite{Singh2016} the twist $\hat{\omega}_0$ is written as $\xi$}
is given by $\hat{\w{\omega}}=\hat{\omega}_0 \w{e_0}$ with
\be \label{omega0rhomu}
\hat{\omega}_0=\smfrac{1}{2}n_{11} = \smfrac{i}{2\sqrt{2}}[(\mu-\overline{\mu})-(\rho-\overline{\rho})].
\ee
At this stage of the construction all variables $I$ of spin-weight 0, such as $\rho, \mu, \epsilon+\overline{\epsilon}, \gamma+\overline{\gamma}, \phi_{00},
\phi_{22}$ and $\phi_{11}$, satisfy $\delta I=0$, while also two algebraic constraints are obtained, namely
\ba
\rho \overline {\mu}-\mu\overline{\rho} &=& 0, \label{rhomurel}\\
\phi_{22}(\rho - \overline{\rho})^2-\phi_{00}(\mu-\overline{\mu})^2 &=& 0.
\ea
GW proceed to show that a rate of rotation in the $(\vm,\vmbar)$ plane exists, such that
\be \label{GWspecrot}
\rho-\overline{\rho}=2 (\epsilon-\overline{\epsilon}),\  \mu-\overline{\mu}=2 (\gamma-\overline{\gamma}),
\ee
implying that
\be
\delta \epsilon = \delta \overline{\epsilon}=\delta \gamma = \delta \overline{\gamma}=0
\ee
and
\be \label{GWalpha}
\delta \alpha = 2 |\alpha|^2+\smfrac{1}{2}F, \ \textrm{ with }\delta{F}=0.
\ee
Next the remaining boost freedom is exploited
%(omitting the special class A solutions, in which $\omega_1=\hat{\omega_1}=0$ and which therefore are of no concern to us, as this condition being preserved under arbitrary boosts)
to make $\rho = e \mu$ and $\epsilon= e \gamma$ ($e=\pm 1$ or $0$),
hereby completely fixing $\w{e_0}$ and $\w{e_1}$, unless $\rho$ or $\mu=0$. I will call the resulting frame the {\em canonical frame}\footnote{obviously $\tilde{\vm}, \tilde{\vmbar}$ remain determined up to constant rotations} $(\tilde{\vk}, \tilde{\vl},\tilde{\vm}, \tilde{\vmbar})$ and indicate with a $\tilde{}$
any of its variables: for example in the canonical frame one has
\be \label{GWrhomu}
\tilde{\rho} = e \tilde{\mu},
\ee
while in any boosted frame (\ref{boosts})
\be
 A^{-1} \rho = e A \mu. \label{boostedrhomu}
\ee
Note that the tetrad choice implying (\ref{GWspecrot}) is identical to the one made by Stewart and Ellis in \cite{Stewart_Ellis_1968}, when they impose on the rotation coefficients the conditions\footnote{both
\cite{GoodeWainwright_1986} and \cite{Stewart_Ellis_1968} construct a canonical tetrad for a general LRS spacetime (i.e.~a general energy-momentum tensor), the only difference arising in the last step,
when Stewart and Ellis choose to make $n_{12}=a_3=0$, while Goode and Wainwright choose $\rho-e\mu=0$}
\be
\Omega_1+\omega_1 = n_{22} = n_{33} = 0 . \label{n22_n33}
\ee
Now {\em any} frame  $(\vk,\vl,\vm,\vmbar)$, in which
$\w{u}=(\vk+\vl)/\sqrt{2}$ is invariant under the action of the isometry group (such as the flow of an imperfect fluid, as considered by Singh et al.~) must be related to the canonical frame by a boost (\ref{boosts}) obeying
\be
\delta A =0.
\ee
Any such boost will leave the relations (\ref{GW_main}--\ref{GWalpha}) invariant, while it will transform (\ref{GWrhomu}) into (\ref{boostedrhomu}), implying by (\ref{omega1rhomu},\ref{omega0rhomu})
\be
\frac{\omega_1}{\hat{\omega}_0}=\frac{e A+ A^{-1}}{e A - A^{-1}}.
\ee
For example the GW relation (\ref{rhomurel}), translated in orthonormal tetrad language, becomes
\be
(\sigma_{11}-\smfrac{2}{3}\theta) \omega_1 -2 a_1 \hat{\omega}_0 =0,
\ee
which is precisely relation (75) of \cite{Singh2016}, where the covariant divergence and the vorticity of $\w{e_1}$ are written respectively as $\phi = -2 a_1$ and $\xi = \hat{\omega}_0$. Using this notation
the conditions (\ref{GWrhomu}) become
\be
\tilde{\sigma}-\smfrac{2}{3}\tilde{\theta} = 0 =\tilde{\xi} \  (e=+1) \textrm{ and } \tilde{\omega}_1 = 0 = \tilde{\phi} \ (e=-1).
\ee
Finally Goode and Wainwright construct coordinates allowing one to write the dual basis $(\tilde{\w{\eta}}^0,\tilde{\w{\eta}}^1,\tilde{\w{\eta}}^2,\tilde{\w{\eta}}^3)$ of the canonical null tetrad $(\tilde{\w{k}},\tilde{\w{\ell}},\tilde{\w{m}},\tilde{\overline{\w{m}}})$ as\footnote{I omit class A, which is LRS-II and has
$\omega_1=\hat{\omega}_0=0$ and I use {\em w} for the GW coordinate {\em u}}
\ba
\tilde{\w{\eta}^0} & = & \frac{\sqrt{2}}{1+e^2} [\ud w + e f(w) \ud v+  e q f(w) e^\psi \Im(\zeta \ud \overline{\zeta})], \label{generalmetric0}\\
\tilde{\w{\eta}^1} & = & \frac{\sqrt{2}}{1+e^2} [-e \ud w + f(w) \ud v+ q f(w) e^\psi \Im(\zeta \ud \overline{\zeta})],\\
\tilde{\w{\eta}^2}  & = & \frac{1}{\sqrt{2}} h(w) e^\psi \ud \zeta, \label{generalmetric}
\ea
with
$f$ and $h$ arbitrary functions of $w$, $q\neq 0$ an arbitrary constant ($q=0$ leading to the non-rotating and non-twisting class A metrics), $e^{-\psi}=1+\smfrac{k}{4}\zeta \overline{\zeta}$ and $e=+1,-1$ or $0$ for classes B1, B2 or B3 respectively. The parameter $k$ represents the Gaussian curvature of the 2-spaces $e^{2\psi} \ud \zeta \ud \overline{\zeta}$ and determines the sign of the function $-F$ appearing in (\ref{GWalpha}), which reduces to $\psi_{\zeta \overline{\zeta}}= -\smfrac{1}{4} k e^{-2\psi}$. Without loss of generality one can put $k=\pm 1$ or $0$, with $k\neq0$ when $e=0$, as otherwise
the spacetime is conformally flat: see (\ref{psi2exp1},\ref{psi2exp2}) below. Note that the coordinates $w$ and $v$ are respectively space-like(time-like) and time-like(space-like) when $e=+1 \ (-1)$ and null when $e=0$. In the latter case (\ref{generalmetric0}--\ref{generalmetric}) is a Kundt spacetime, with $k$ determining the Gaussian curvature of the $w=constant$ wave fronts.
With respect to an orthonormal tetrad $(\w{e_0},\w{e_1},\w{e_2},\w{e_3})$ defined by (\ref{e0e1def},\ref{mmbardef}) and with $\vk, \vl$ related to the canonical
null vectors $\tilde{\vk}, \tilde{\vl}$ by a boost (\ref{boosts}) which is invariant under the full isometry group ($A=A(w)$), the non-vanishing kinematic quantities of an `imperfect fluid flow' $\w{u}=\w{e_0}$
are then given by
\ba
 \dot{u}_1 &=&  \frac{(f A_{+})'}{2 f},\label{u_acc}\\
 \omega_1 &=& \frac{ 2 q f A_{+}}{(e^2+1) h^2},\label{u_omega}\\
 \theta &=&  - \frac{(f h^2 A_{-})'}{2 f h^2},\label{u_theta}\\
 \sigma_{11} &=& 2 \sigma_{22}=2 \sigma_{33} = - \frac{h}{3 f}(\frac{f A_{-}}{h})' ,\label{u_sigma}
\ea
while the twist and the expansion of $\w{e_1}$ are given by
\ba
 \hat{\omega}_0 &=& \frac{ 2 q f A_{-}}{(e^2+1) h^2},\label{e_omega}\\
 \hat{\theta} &=&  \frac{(f h^2 A_{+})'}{2 f h^2},\label{e_theta}
\ea
($A_{\pm} = A e \pm A^{-1}$ and a dash indicates a derivative with respect to $w$).
Note that the arbitrary function $A(w)$ does not enter the metric $2 \w{\eta}^0\w{\eta}^1-2\w{\eta}^2\w{\eta}^3$, but only the boosted tetrad via the relations (\ref{boosts}).

Also note that in classes B1 and B2, while the `canonical flow' (determined by $A=1$) is always non-rotating or non-expanding, as
well as non-rotating or non-twisting ($\tilde{\omega}_1 \tilde{\theta}=\tilde{\omega}_1 \hat{\tilde{\omega}}_0 = 0$), this is not necessarily so for a boosted flow.

The $\Psi_2$ component of the Weyl spinor w.r.t.~the tetrad $(\vk,\vl,\vm,\vmbar)$ is given by\footnote{(\ref{psi2exp2}) implies that class B3 is purely electric, in agreement with \cite{LozanovskiCarminati2003}}
\ba
\Re \Psi_2 &=& \frac{2 e f^{-1}}{3(3e^2+1)} \{[h(\frac{f}{h})']'- 4 q^2 f^3 h^{-4}\} -\frac{k}{6h^2} , \label{psi2exp1}\\
\Im \Psi_2 &=& \frac{4 e q}{3e^2+1} h^{-1} (\frac{f}{h})' \label{psi2exp2}
\ea
and the components of the energy-momentum tensor (\ref{energymomentum}) with respect to $(\w{e_0},\w{e_1},\w{e_2},\w{e_3})$ by
\begin{enumerate}[i)]
\item (when $e=\pm 1$)
\ba \label{en_e_nonzero}
T^{00} &=& -\smfrac{1}{2} A_{+}^2 \frac{h''}{h} +\smfrac{1}{2} A_{-}^2 \frac{f'}{f}\frac{h'}{h}-e \frac{h'^2}{h^2} \nonumber \\
&& +\frac{q^2}{4}f^2 h^{-4}(3 A_{+}^2-A_{-}^2)+kh^{-2},\\
T^{01} &=& \smfrac{1}{2} (A^2-A^{-2}) f h^{-1} [q^2 f h^{-3} - (\frac{h'}{f})'],\\
T^{11} &=& -\smfrac{1}{2} A_{-}^2 \frac{h''}{h} +\smfrac{1}{2} A_{+}^2 \frac{f'}{f}\frac{h'}{h}+ e \frac{h'^2}{h^2} \nonumber \\
&& +\frac{q^2}{4}f^2 h^{-4}(3 A_{-}^2-A_{+}^2)-kh^{-2},\\
T^{22} &=& T^{33}= e[\frac{(f h)''}{f h}-\frac{f'}{f}\frac{h'}{h}]+e q^2 f^2 h^{-4},
\ea
\item (when $e=0$ and hence $k\neq 0$)
\ba
%\w{T}= 2 A^2 [k h^{-2} (\w{e_0}\otimes \w{e_0}-\w{e_1}\otimes \w{e_1})+(\frac{q^2f^2}{h^4}-\frac{h''}{h}+\frac{h'}{h}\frac{f'}{f})(\w{e_0}-\w{e_1})\otimes (\w{e_0}-\w{e_1})].
T^{00} &=& \smfrac{1}{2} A^{-2}(\frac{4 q^2 f^2}{h^4}-\frac{h''}{h}+\frac{h'}{h}\frac{f'}{f}) +kh^{-2}, \label{en_e_iszero} \\
T^{01} &=& -\smfrac{1}{2} A^{-2}(\frac{4 q^2 f^2}{h^4}-\frac{h''}{h}+\frac{h'}{h}\frac{f'}{f}),\\
T^{11} &=& \smfrac{1}{2} A^{-2}(\frac{4 q^2 f^2}{h^4}-\frac{h''}{h}+\frac{h'}{h}\frac{f'}{f}) -kh^{-2},\\
T^{22} &=& T^{33}= 0. \label{en_e_iszero2233}
\ea
\end{enumerate}
Note that (\ref{psi2exp1},\ref{psi2exp2}) imply that the spacetime is purely (Weyl-) electric when $e=0$ or $f/h=constant$ and purely (Weyl-) magnetic when $s=f/h$ is a solution of
\be
\frac{\ud^2 s}{\ud u^2} - e k  s -4 q^2 s^3=0,\ \  (u = \int h^{-1} \ud w \textrm{ and } e=\pm 1),
\ee
in agreement with the results obtained by Lozanovski and Carminati \cite{LozanovskiCarminati2003}.

\section{The self-similar case}
  We now find out  which of the B1, B2 or B3 spacetimes above admits a homothetic Killing vector and hence corresponds to one of the particular self-similar cases discussed in \cite{Singh2016, Singh2017}.
  The question is facilitated by the fact that each of the spacetimes (\ref{generalmetric0}--\ref{generalmetric}) is hypersurface-homogeneous, such that the existence of a proper homothetic Killing vector implies the constancy of
  all dimensionless Cartan invariants in the Cartan-Karlhede classification \cite{Kramer}. Using the fact that $\tilde{\w{e_0}},\tilde{\w{e_1}}$ are invariantly defined, one obtains, after substituting $A=1$ in
  for example (\ref{u_acc},\ref{u_omega},\ref{e_theta}) when $e=+1$ or $0$ and in (\ref{u_theta},\ref{u_sigma},\ref{e_omega}) when $e=-1$,  that $f= f_0 h^n$ ($f_0,n$  constants).
 Substituting this in the expression (\ref{psi2exp1},\ref{psi2exp2}) and
expressing that $\Psi_2/\omega_1^2$ (or $\Psi_2/{\hat{\omega}_0}^2$) are constants, only one of the following cases survives:
\begin{enumerate}[i)]
\item ($f_0,h_0,c$ constants, with $c$ possibly $0$)
\be
f(w)= f_0 h_0^2 e^{2 c w} \textrm{ and } h(w)= h_0 e^{c w}
\ee
\item ($f_0, h_0, c$ constants, with $c=1$ when $k\neq 0$)
\be
f(w)= f_0 h_0^2 w^{2 c -1} \textrm{ and } h(w) = h_0 w^c .\label{selfsimcase}
\ee
\end{enumerate}
We dismiss case (i), as it leads to homogeneous Petrov type D spacetimes, for which it is known that they do not admit proper homothetic Killing vectors \cite{Koutras_Skea_1998}.
On the other hand, case (ii) does indeed result in self-similar spacetimes, with a homothetic Killing vector being given by
 \be
 \w{H}= w \partial_w +(1-c)(2 v \partial_v + \zeta \partial_{\zeta}+\overline{\zeta} \partial_{\overline{\zeta}}).
 \ee
With respect to the boosted tetrad the components of the energy-momentum tensor and the real and imaginary parts of the Weyl spinor become now\\

\noindent
a)  when $e=\pm 1$
\ba
w^2 T^{00} &=& \frac{1}{2} (q^2 f_0^2+ c^2)A_{+}^2 \nonumber \\
&& + e ( q^2 f_0^2-5 c^2+2 c)+k h_0^{-2},\\
w^2 T^{01} &=& \frac{1}{2} (A^2-A^{-2}) (q^2 f_0^2+c^2) ,\\
w^2 T^{11} &=& \frac{1}{2} (q^2 f_0^2+ c^2)A_{-}^2 \nonumber \\
&& - e ( q^2 f_0^2-5 c^2+2 c)-k h_0^{-2},\\
w^2 T^{22} &=& w^2 T^{33} = e( q^2 f_0^2+7 c^2-8 c + 2),\\
w^2 \Re \Psi_2 &=& \frac{e}{3} (c-1)^2-\frac{2}{3}q^2 f_0^2 -\frac{k}{6}h_0^{-2},\\
w^2 \Im \Psi_2 &=& e q f_0(c-1).
\ea
and\\

\noindent
b) when $e=0$ (hence $k\neq 0$ and $c=1$)
\ba
w^2 T^{00} &=& \smfrac{1}{2} A^{-2} (4 q^2 f_0^2+1)+k h_0^{-2},\\
w^2 T^{01} &=& -\smfrac{1}{2} A^{-2} (4 q^2 f_0^2+1) ,\\
w^2 T^{11} &=& \smfrac{1}{2} A^{-2} (4 q^2 f_0^2+1)-k h_0^{-2},\\
w^2 T^{22} &=& w^2 T^{33} = 0,\\
w^2 \Psi_2 &=& -\frac{k}{6}h_0^{-2}.
\ea 

Note that, when $k\neq 0$, these spacetimes are all purely (Weyl-) electric, while for $k=0$ purely (Weyl-) magnetic spacetimes occur when $q=\frac{c-1}{\sqrt{2}f_0}$ and hence 
\be 
\Psi_2= i \frac{e (c-1)^2}{\sqrt{2}}w^{-2}.
\ee

\section{Discussion}
I have shown that the general solution for the most general class of LRS spacetimes, as claimed in \cite{Singh2016, Singh2017}, contains a much larger set of solutions than the self-similar ones. In fact, for the non-conformally
flat case, any of the spacetimes (\ref{generalmetric}) will do, provided one chooses the vector field $\w{u}$ in the plane of the principal null
directions of the Weyl tensor to be invariant under the given isometry group. These vector fields $\w{u}$ can be obtained by boosting the canonical GW frame, with the boost scalar $A$ being an arbitrary function of
$w$. \\
The question remains whether the resulting energy-momentum tensors are physically relevant, in the sense that they satisfy for example the weak energy condition. Obviously it suffices
to answer this question in the canonical frame itself, i.e.~by substituting $A=1$ in the expressions (\ref{en_e_nonzero}--\ref{en_e_iszero2233}).

When $e=0$ it suffices to express that for $|x| \leqslant 1$
\be
x^2 T^{11}+ 2 x T^{01}+T^{00} \equiv (1-x)[a(1+x)+b(1-x)] \geqslant 0
\ee
($a=\smfrac{1}{2} (\frac{4 q^2 f^2}{h^4}-\frac{h''}{h}+\frac{h'}{h}\frac{f'}{f})$
and $b=kh^{-2}$). The weak energy condition will therefore be satisfied if and only if
\be
k=+1 \textrm { and } \frac{4 q^2f^2}{h^4}-\frac{h''}{h}+\frac{h'}{h}\frac{f'}{f} > 0,
\ee
a condition which in the self-similar case (\ref{selfsimcase}) will hold for arbitrary values of $q$ and $f_0$. In this case ($e=0,c=1,k=\pm 1$) the conformally related spacetime with metric 
\be 
\ud s^2=w^2 (2 \w{\eta}^0 \w{\eta}^1-2 \w{\eta}^2\w{\eta}^3)
\ee
 is homogeneous and admits two null Killing vectors ($\partial_v$ and $\partial_w$), both parallel to the principal null directions of the Weyl tensor and with $\partial_v$ being covariantly constant. No physical interpretation of these Kundt spacetimes, nor of the conformally related homogeneous ones, is available.

On the other hand when $e=\pm 1$ and $A=1$ we have $T^{01}=0$ and the weak energy condition
requires $T^{00}, T^{00}+T^{11},T^{00}+T^{22}$ to be positive, which in the self-similar case becomes\\
\begin{enumerate}[a)]
\item when $e=+1$ and $k=0$,
\be
f_0^2 q^2 > \textrm{max}(c^2-\smfrac{2}{3}c,-c^2+\smfrac{3}{2} c-\smfrac{1}{2}),
\ee
when $e=+1$ and $k\neq 0, c=1$,
\be
 f_0^2q^2 >  \textrm{max}(\smfrac{1}{3}-\smfrac{k}{3}h_0^{-2},-\smfrac{k}{4}h_0^{-2})
\ee
\item when $e=-1$ and $k=0$,
\be
f_0^2q^2 < \textrm{min}(5 c^2-2 c,-2(c^2-3 c+1)),
\ee
when $e=-1$ and $k\neq 0, c=1$,
\be f_0^2q^2 <  \textrm{min}(1+\smfrac{k}{2}h_0^{-2},3+k h_0^{-2}).
\ee
\end{enumerate}
Again no physical interpretation of these spacetimes is available.

\begin{acknowledgments}
All calculations were done with the aid of the Maple packages DifferentialGeometry and GRTensorIII.
\end{acknowledgments}

\end{document}